\documentclass[conference]{IEEEtran}

\usepackage{cite}

\usepackage{amsmath,amssymb,amsfonts}
\usepackage{mathtools}

\usepackage{algorithm}
\usepackage{algpseudocode}

\usepackage{graphicx}
\usepackage{textcomp}
\usepackage[normalem]{ulem}
\usepackage{color}
\usepackage{xcolor}

\usepackage{array}
\usepackage{booktabs}
\usepackage{colortbl}
\usepackage{multirow}

\def\BibTeX{{\rm B\kern-.05em{\sc i\kern-.025em b}\kern-.08em
    T\kern-.1667em\lower.7ex\hbox{E}\kern-.125emX}}

\title{ARAS: An Adaptive Low-Cost ReRAM-Based Accelerator for DNNs\vspace{-.4cm}}

\author{
    \IEEEauthorblockN{Mohammad Sabri, Marc Riera, Antonio González}
    \IEEEauthorblockA{Universitat Polit\`{e}cnica de Catalunya (UPC)
    \\mohammad.sabri@upc.edu, marc.riera.villanueva@upc.edu, antonio@ac.upc.edu}
}

\begin{document}
\maketitle
\thispagestyle{plain}
\pagestyle{plain}

\begin{abstract}
Processing Using Memory (PUM) accelerators have the potential to perform Deep Neural Network (DNN) inference by using arrays of memory cells as computation engines. Among various memory technologies, ReRAM crossbars show promising performance in computing dot-product operations in the analog domain. Nevertheless, the expensive writing procedure of ReRAM cells has led researchers to design accelerators whose crossbars have enough capacity to store the full DNN. Given the tremendous and continuous increase in DNN model sizes, this approach is unfeasible for some networks, or inefficient due to the huge hardware requirements. Those accelerators lack the flexibility to adapt to any given DNN model, facing an \textit{adaptability} challenge.

To address this issue we introduce ARAS, a cost-effective ReRAM-based accelerator that employs a smart scheduler to adapt different DNNs to the resource-limited hardware. ARAS also overlaps the computation of a layer with the weight writing of several layers to mitigate the high writing latency of ReRAM. Furthermore, ARAS introduces three optimizations aimed at reducing the energy overheads of writing in ReRAM. Our key optimization capitalizes on the observation that DNN weights can be re-encoded to augment their similarity between layers, increasing the amount of bitwise values that are equal or similar when overwriting ReRAM cells and, hence, reducing the amount of energy required to update the cells. Overall, ARAS greatly reduces the ReRAM writing activity. We evaluate ARAS on a popular set of DNNs. ARAS provides up to $2.2\times$ speedup and $45\%$ energy savings over a baseline PUM accelerator without any optimization. Compared to a TPU-like accelerator, ARAS provides up to $1.5\times$ speedup and $61\%$ energy savings.
\end{abstract}


\section{Introduction}\label{s:intro}
Processing Using Memory (PUM) is a promising approach to mitigate the high computational and energy costs associated with the inference of Deep Neural Network (DNN) models. Various Non-Volatile Memory (NVM) technologies, including Phase-Change Memory (PCM)~\cite{PCM}, Spin-Transfer Torque Magnetic RAM (STT-MRAM)~\cite{STT}, and Resistive-RAM (ReRAM)~\cite{ResiRCA}, can efficiently implement the pivotal dot-product operations required for DNNs in the analog domain. Among these, ReRAM technology stands out with its lower read latency and higher density, making it a suitable candidate for designing DNN accelerators. Therefore, a growing body of work~\cite{PRIME, sparse_ReRAM, CASCADE} is exploring ReRAM crossbars due to their dense and efficient analog computation capabilities.

Despite the promising features of ReRAM, its high write latency and write energy consumption pose severe limitations in designing efficient DNN accelerators. As a result, prior works have proposed to process multiple DNN inferences on a pipelined mode to mitigate the writing overheads of ReRAM~\cite{RAPIDNN, FORMS, ISAAC, TIMELY, AtomLayer, NVMExplorer, RAELLA, Neurosim_mapping, PUMA}. Besides, these accelerators are designed under the assumption of having enough ReRAM crossbars to accommodate all network weights, that is, the writing of DNN weights is only done once at the beginning of the first inference. While this approach addresses the inefficiency of ReRAM writing, it also introduces a significant issue that impedes the future use of ReRAM-based accelerators as a prominent compute platform.


The rapid growth in the complexity of DNNs has led to a dramatic increase in network model sizes every year~\cite{DNA-TEQ, Qeihan}. To illustrate it, AlexNet~\cite{Alex-net} had 144MB of parameters in 2012, while the BERT-Large model~\cite{BERT}, introduced in 2018, comprises a substantial 1.3GB of parameters, and Chat-GPT 3.5~\cite{chatGPT}, released in 2023, pushed the boundaries to 800GB. Current ReRAM-based accelerators face an \textit{adaptability} issue because the crossbars they employ today may not be sufficient for handling larger networks in the future. In addition, scaling the resources of an accelerator to the biggest DNN is not the most effective solution in terms of area and power. Similar to other systems, such as TPUs and GPUs, ReRAM-based accelerators should be flexible and capable of efficiently executing any type of DNN, regardless of its size, using limited resources. Most accelerators are typically initialized for a specific DNN model and lack mechanisms for efficiently updating weights to accommodate different models at runtime. This issue hinders the widespread use of ReRAM-based accelerators, especially in server environments where there are continuous requests to perform inference for different DNN models. Hence, a practical ReRAM-based accelerator must address this \textit{adaptability} challenge to ease its adoption.

In this paper, we present ARAS, an Adaptive low-cost ReRAM-based Accelerator for DNN inference. ARAS main novelty is an offline scheduler and its corresponding hardware support that includes efficient mechanisms for programming DNN weights into a limited-size accelerator at runtime. The execution of a DNN model involves various tasks, such as data transfers, weight updating, and dot-product computations, among others, which must be properly scheduled in the accelerator to efficiently manage its resources. To the best of our knowledge, this is the first work aimed at devising a scheduler capable of optimizing the utilization of a ReRAM compute engine. ARAS performs one DNN inference at a time and employs a new execution approach, supporting a broad range of different layers and DNN models. ARAS prioritizes the allocation of available ReRAM crossbars for the concurrent weight updating of upcoming layer(s). It then conducts the associated computations sequentially, layer-by-layer, due to data dependencies across layers, iterating until all required computations for the network are completed. This approach enables the accelerator to methodically process large networks even with constrained resources, maximizing efficiency and resource utilization.


The weight updating task has a significant impact on both performance and energy consumption due to the high writing latency and write energy costs of ReRAM. Consequently, ARAS incorporates a set of optimizations to counter the inherent inefficiencies of ReRAM writes. To mitigate these overheads, ARAS employs a scheduling strategy that overlaps the latency of computing a layer with the latency of writing weights for the next layer(s). In addition, the accelerator includes a multi-banked global buffer with banks of different sizes whose purpose is to reduce static energy consumption. Furthermore, ARAS employs an adaptive weight replication scheme to improve the execution of long compute latency layers such as some convolutions. This scheme is designed to reduce latency in critical layers by performing computations for multiple convolution steps simultaneously. Our approach requires to search for a replication factor that finds the best trade-off between the latency reduction and the cost of replicating the weights. Finally, ARAS leverages the concept of partial weight reuse when updating ReRAM cells to write new layer weights. Given that each weight is represented by multiple ReRAM cells, weight reuse refers to the equality or similarity of the current value of a cell and the new value to be overwritten in it. ARAS aims to increase partial reuse across weights between consecutive layers, avoiding ReRAM cell updates for certain bits of weights, and saving the corresponding energy required for writing.

To summarize, this work focuses on designing ReRAM accelerators that are adaptable to efficiently execute any present and future DNN, regardless of its model size. The main contributions are:


\begin{itemize}

\item We have designed an adaptive accelerator that enables concurrent neural computation and network weight updates. ARAS incorporates an offline scheduler that orchestrates the execution of layers, prioritizing the writing of new weights as soon as ReRAM crossbars finish their assigned computations.



\item We employ a heterogeneous multi-banking scheme that partitions the on-chip memory responsible for storing the DNN's input and output activations. This scheme aims to select the smallest possible banks based on the capacity required to store the activations of each layer, reducing static energy by 3\% on average.

\item We introduce an adaptive weight replication scheme to enhance the accelerator's throughput by concurrently computing multiple convolution steps. ARAS determines the best replication factor taking into account the overheads of additional weight writes and the speedup that can be achieved in each layer, allocating ReRAM crossbars based on the computation latency and available crossbars.

\item We develop a novel partial weight reuse scheme with the goal of maximizing the partial bit-level similarity among weights of different layers. This scheme allows us to skip the updating of cells when the bits to be overwritten are identical to the new ones, thus reducing the number of ReRAM cells that are updated. Our key observation reveals that aligning the mean of layers' weights to a common value enables efficient reuse of a substantial portion of the most significant bits, reducing ReRAM writing activity by 17\% on average.

\item We evaluate ARAS on five representative DNNs. On Average, ARAS's novel scheduler and its associated optimizations improve performance and reduce energy by \(1.5\times\) and 28\%, respectively. Compared to a TPU-like accelerator, ARAS achieves, on average, \(1.2\times\) speedup and 33\% reduction in energy consumption.

\end{itemize}


\section{Background and Related Work}\label{s:background}

\subsection{ReRAM Cell and Crossbar Architecture}\label{subs:ReRAM}
\begin{figure}[t!]
    \centering
    \includegraphics[width=0.7\columnwidth]{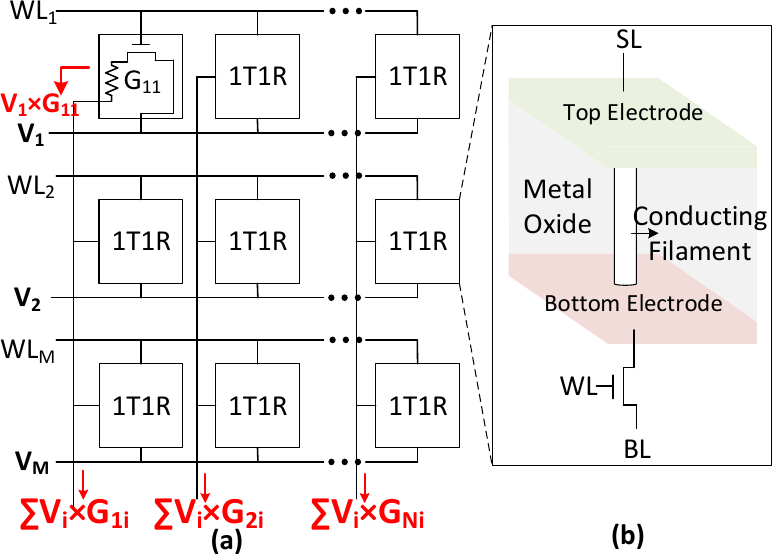}
    \vskip -0.10in
    \caption{Crossbar architecture (a) and analog operation with ReRAM cells (b).}
    \vskip -0.15in
    \label{fig:CrossBars}
\end{figure}

ReRAM cells are two-terminal devices (see Figure~\ref{fig:CrossBars}(b)) able to represent values through their analog multilevel conductance states. In DNN accelerators, ReRAM cells are used to store the network weights. The process of encoding or writing these weights (weight updates) relies on transitions between various conductance states in the ReRAM cells, typically initiated by electrical inputs. The physical mechanism governing these conductance transitions may vary for different types of ReRAM insulators~\cite{insulator, insulator2, Park_2013, Woo_2016}. Generally, the conductance of ReRAM cells can be augmented or reduced using positive and negative programming voltage pulses, denoted as weight increase and weight decrease, respectively.


The most compact and simple structure for creating a DNN weight matrix with ReRAM cells is the crossbar array configuration, in which each 1T1R cell is positioned at every intersection point. In addition, crossbar structures offer the advantage of achieving high integration density~\cite{NeuroSim_writing}. ReRAM crossbars serve a dual purpose by storing the weights and executing the dot-product operations between inputs and weights. As depicted in Figure~\ref{fig:CrossBars}, when the input vector is encoded using read voltage signals, the weighted sum operation (equivalent to matrix-vector multiplication) can be executed in parallel using the weights stored in the crossbar. The results of this weighted sum are obtained at the end of each column, represented as analog currents, and post-processed by additional peripherals.

\subsection{ReRAM Writing}\label{subs:Writing}
\begin{figure}[t!]
    \centering
    \includegraphics[width=0.70\columnwidth]{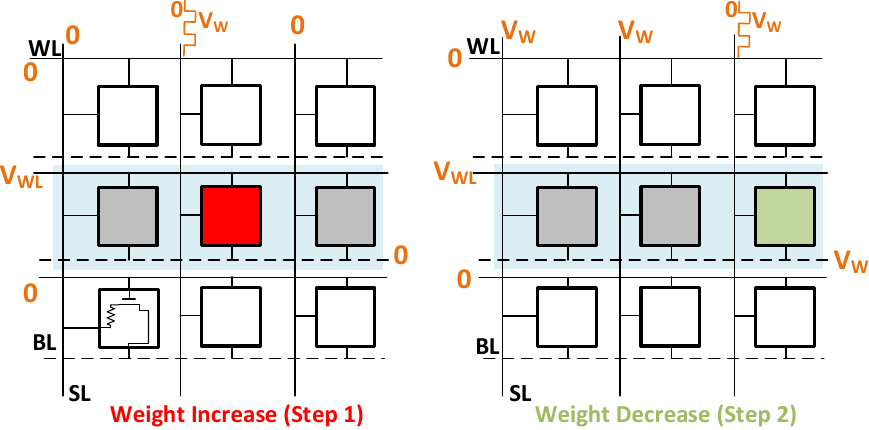}
    \vskip -0.10in
    \caption{Example of a Row-by-Row writing scheme. The gray cell keeps its value in both steps, while the red and green cell values are modified.}
    \label{fig:WeightUpdating}
    \vskip -0.15in
\end{figure}

Gao et al.~\cite{Gao_2015} introduced a fully parallel writing scheme to update weights in ReRAM crossbars. However, programming all the cells in the array simultaneously might demand more peak power than the peripheral circuits can supply. As a result, a row-by-row writing scheme~\cite{NeuroSim_writing} is typically employed for the weight update process, as shown in Figure~\ref{fig:WeightUpdating}. The weight increase and decrease operations require distinct programming voltage polarities, so the weight update process is divided into two steps, each with its specific voltage polarity for every row. In each step, the SLs (Source Lines) provide voltage pulses or constant voltages (if no update) to modify each selected cell, while the BL (Bit Line) provides the required polarity for that particular step. Ideally, selecting the entire row at once is the most effective approach to ensure maximum parallelism. However, it is possible that only a portion of a row is selected at a time if the write circuitry cannot handle the substantial write current required for the entire row. It is important to note that in ReRAM cells, the update of a cell's value is achieved by applying specific pulses through the associated SL, and the number of pulses required is determined by the current cell value and the desired target value. Consequently, in a crossbar array, each SL (crossbar column) is equipped with its independent driver to deliver distinct pulses.

Current commercial ReRAM storage chips are primarily designed for single or few-time writes, resulting in relatively low endurance compared to other memory technologies. This underscores a notable gap in practical ReRAM cell development that is well-suited for PUM applications. However, there is a growing body of research that developed different insulators to enhance endurance cycles to levels around $10^{11}$ and $10^{12}$~\cite{Endurance_1, Endurance_2, Endurance_3, Endurance_4, Endurance_6}. Additionally, some research~\cite{Endurance_5} indicates the potential for ReRAM endurance of up to $10^{15}$, although achieving it may be a long-term goal.

\subsection{DNN Mapping}\label{subs:Mapping}
\begin{figure}[t!]
    \centering
    \includegraphics[width=0.8\columnwidth]{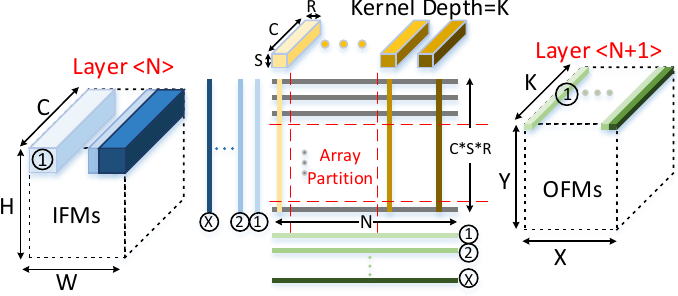}
    \vskip -0.15in
    \caption{Conventional mapping of a CONV layer in ReRAM crossbars.}
    \label{fig:Conventional_Mapping}
    \vskip -0.20in
\end{figure}

DNN mapping~\cite{Neurosim_mapping, Training_map, PattPIM} has a significant impact on the performance and energy consumption of ReRAM-based accelerators. ARAS employs a conventional scheme, shown in Figure~\ref{fig:Conventional_Mapping}, where each kernel of a CONV layer is unrolled into a single column of the ReRAM crossbar, which may be further partitioned into multiple arrays depending on the number of weights and the number of bits per weight. Then, a window of activations is also unrolled and applied to all the kernels in a single step. In the next step, the window slides over, and the new activations are streamed bit-serially to the crossbar again. Similarly, a FC layer is a special case of CONV layer.


\subsection{Weight Replication}\label{subs:Replication}
Weight replication is a key technique used in prior ReRAM-based accelerators~\cite{PipeLayer, TIMELY}. This technique is mainly used in Convolutional Neural Networks (CNNs) and aims to boost throughput by replicating the weights of convolution layers in ReRAM crossbars, computing multiple activation windows simultaneously. The advantages of replication are particularly pronounced in the initial layers of CNNs, which have relatively small weight matrices but a vast number of activation windows. Therefore, for these layers, given that the kernel sets are small, the replication costs are modest, and the improvement in throughput is substantial. In previous works~\cite{RAELLA, Neurosim_mapping}, replication follows a greedy scheme: as long as there are available resources, the layer with the lowest throughput is chosen for replication. However, it is crucial to balance the replication factor for each layer. An inappropriate replication factor could adversely impact performance and energy efficiency. This is because it may lead to additional writes in accelerators where resources are limited, including both ReRAM crossbars for weight storage and on-chip buffers for partial results.

\section{ARAS Accelerator}\label{s:ARAS Architecture}

\subsection{Architecture}\label{subs:Architecture Components}
\begin{figure*}[t!]
    \centering
    \includegraphics[width=0.70\textwidth]{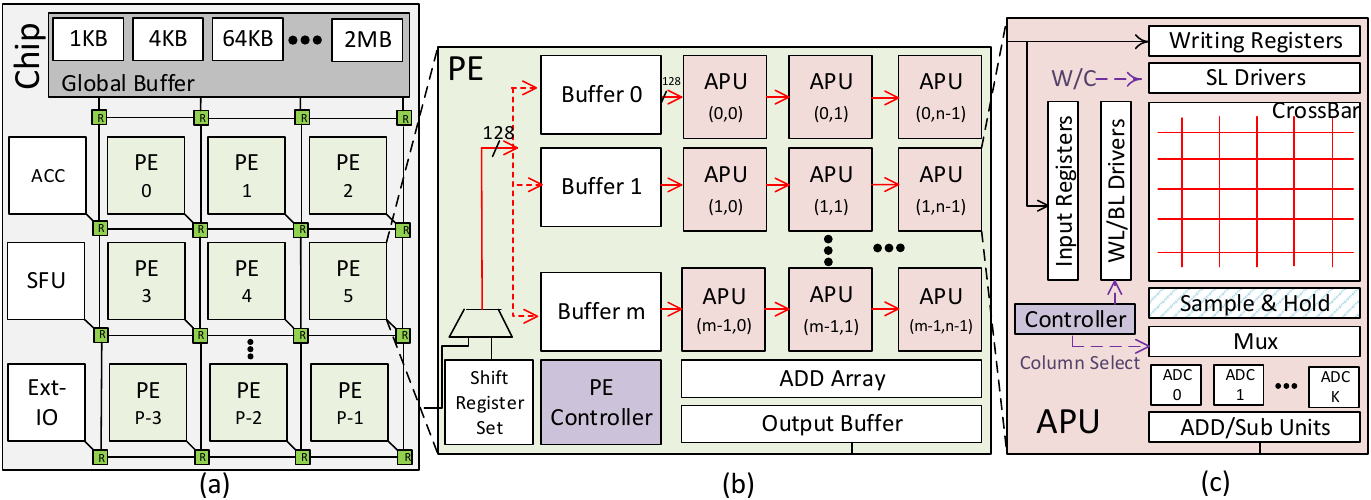}
    \vskip -0.10in
    \caption{Architecture of the ARAS accelerator including the organization of: (a) Chip, (b) Processing Element (PE), and (c) Analog Processing Unit (APU).}
    \vskip -0.15in
    \label{fig:Top_view}
\end{figure*}

In this section, we present the ARAS architecture. Previous ReRAM-based accelerators~\cite{Neurosim_mapping, ISAAC, ReDy} assume that all DNN weights are pre-stored in the ReRAM crossbars. These designs often employ a deep pipeline capable of executing multiple DNN inferences concurrently. However, such an approach demands substantial amount of resources, including both ReRAM crossbars and on-chip SRAM buffers, which is not the most effective solution in the long term. In contrast, ARAS operates with limited resources, adopting a layer-by-layer execution that overlaps the computation of a given layer with the procedure of writing weights for subsequent layers. Figure~\ref{fig:Top_view} shows a block diagram top-down view of the architecture of the ARAS accelerator.


Figure~\ref{fig:Top_view}(a) depicts a high-level schematic of the chip architecture. A single chip comprises an External IO Interface (Ext-IO), multiple PEs, a Special Function Unit (SFU), Accumulation Units (ACC), and a Global Buffer (Gbuffer). The Ext-IO is used to communicate with Main Memory (MM) to load the network weights and inputs and store its final outputs. The ACC units are responsible for aggregating all the partial results of the neural operations for a given layer. These partial results may be generated by various PEs when a layer occupies multiple PEs due to its size. SFU performs transitional functions like pooling, non-linear activations (i.e. sigmoid or ReLU), and normalization, to support the full range of computations required for state-of-the-art DNNs. These components work in concert to enable efficient end-to-end DNN inference execution and support a broad spectrum of neural network operations.

The Global Buffer (Gbuffer) serves as a storage unit for intermediate activations generated during the execution of a layer. To optimize on-chip memory utilization, the Gbuffer is structured with banks of varying sizes. This design allows for a variable allocation of memory banks based on the size of the layer being processed. Consequently, inefficiency in memory allocation is mitigated. This adaptive approach contributes to the overall energy efficiency of ARAS since those banks that do not include essential information are power-gated.

Figure~\ref{fig:Top_view}(b) presents the structure of a PE, which contains $m \times n$ APUs, $m$ buffers to store either input activations to process or weights to write, an output buffer to store partial sums, $n$ accumulation modules to add the partial sums from different APUs, a set of shift registers to serialize the activations, and a multiplexer to switch between writing and computing phases. Finally, Figure~\ref{fig:Top_view}(c) shows the main components of an APU including a crossbar array to store synaptic weights that is built based on ReRAM cells (1T1R). Each APU also includes an input register to store activations, a writing register to store weight deltas, a WL/BL switch matrix to drive the wordlines and bitlines, an SL switch matrix to drive the sourcelines, an analog multiplexer, a shared pool of ADCs, and functional units to accumulate and shift the partial results from different BLs and iterations. Section~\ref{s:background} provides more details on the analog dot-product operation with ReRAM crossbars and the writing of ReRAM cells.


\subsection{Execution Dataflow}\label{subs:computation/Weight Updating Procedures}
\begin{figure}[t!]
    \centering
    \includegraphics[width=1.0\columnwidth]{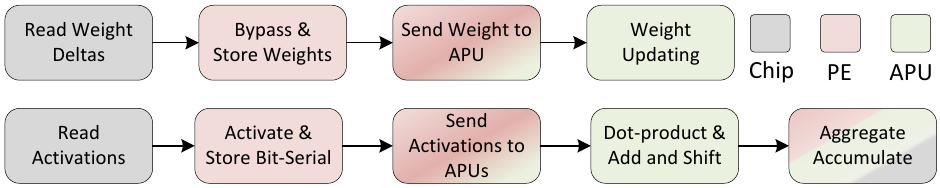}
    \vskip -0.10in
    \caption{ARAS execution dataflow. The top flowchart shows the weight writing procedure while the bottom corresponds to the computation.}
    \label{fig:Dataflow}
    \vskip -0.20in
\end{figure}

In this section, we explain the dataflow followed by the ARAS accelerator. Figure~\ref{fig:Dataflow} illustrates the procedures for writing weights and performing computations. Each level of the top-down architecture is represented by a different color.

\emph{Writing Procedure Dataflow:} The top of Figure~\ref{fig:Dataflow} shows the weight writing flowchart for a given ReRAM crossbar row. Initially, the deltas of the weights of a row are fetched from main memory. As described in Section~\ref{subs:Writing}, the number of writing pulses required to update the value of multi-level ReRAM cells depends on the current value of the cell and its next coming value. The difference between current and new values is referred to as weight deltas, calculated offline, and directly proportional to the energy of writing weights. Then, these deltas are transferred through a NoC to the corresponding PE, where the shift registers are bypassed to store the deltas directly in the destination buffer. Next, the target APU reads deltas from the corresponding buffer and stores them in its Writing Registers. Finally, the APU drivers are configured to perform weight updating by decreasing and increasing the cell values in two phases (Figure~\ref{fig:WeightUpdating}). Note that the concurrent writing of all cells in a crossbar row entails that the row's writing latency is determined by the cells that exhibit the highest latency in each phase, that is, the slowest cells in the row are the bottleneck of the writing operation. Besides, ARAS considers the main memory bandwidth as a potential bottleneck when many APUs are concurrently updating the weights, limiting the simultaneous writing of APUs. In this work, we focus on reducing the magnitude of the weight deltas to improve energy consumption, so ReRAM writing latency remains an open challenge for future work.


\emph{Computation Procedure Dataflow:} The bottom of Figure~\ref{fig:Dataflow} summarizes the compute dataflow for a given convolutional window. First, activations are fetched and distributed to the PEs based on the corresponding location of the written weights, as shown in Figure~\ref{fig:Conventional_Mapping}, where all the weights stored in a given row of APUs belong to a single layer. Then, activations undergo a serialization process within the PEs, the PE controller activates the associated buffer, and activations are streamed and stored bit-serially into the buffer. Unlike the work in \cite{FORMS}, the shift register set, which is in charge of serializing activations, is shared for different rows of APUs in a PE to mitigate area overhead. Therefore, to avoid harming latency, the number of rows of APUs should be limited to the throughput of the shift register set. Similar to previous works~\cite{ISAAC, PRIME}, ARAS iterates the activations bit by bit, then performs dot-product operations for each activation bit, and finally shifts and accumulates with the previous iterations. Therefore, the total number of iterations depends on the activation's precision. In the last iteration, APUs return their part of the neural computation, and the final results related to each kernel/filter is computed by accumulating all APUs' partial results. According to the kernels' dimension and layer mapping, partial results accumulation can be performed inside PEs or in the chip accumulator. Finally, the outputs of a window passing through activation and pooling functions before stored back to the Global Buffer, and the accelerator is configured to compute the next window.


Note that ARAS takes an efficient approach by allowing PEs to write weights from different layers, but computing one layer at a time. This strategy, although more complex in terms of control and scheduling, optimizes resource usage, resulting in improved throughput and energy efficiency.

\section{ARAS Scheduler}\label{s:ARAS_Scheduler}

In this section, we present the proposed offline scheduler, which plays a key role in managing both computation and weight writing tasks, by efficiently assigning the resources needed for each task. Our scheduler follows a systematic layer-by-layer computation approach, that is, as soon as one DNN layer's computation is completed, the next layer can start processing. This approach allows ARAS to overlap the computations of a layer and the writing of weights of the following layers. Furthermore, it leverages the concurrent updating of multiple ReRAM crossbars, aiming to maximize the utilization of the resources and minimize the latency associated with the writing process.

\begin{figure}[t!]
    \centering
    \includegraphics[width=1.00\columnwidth]{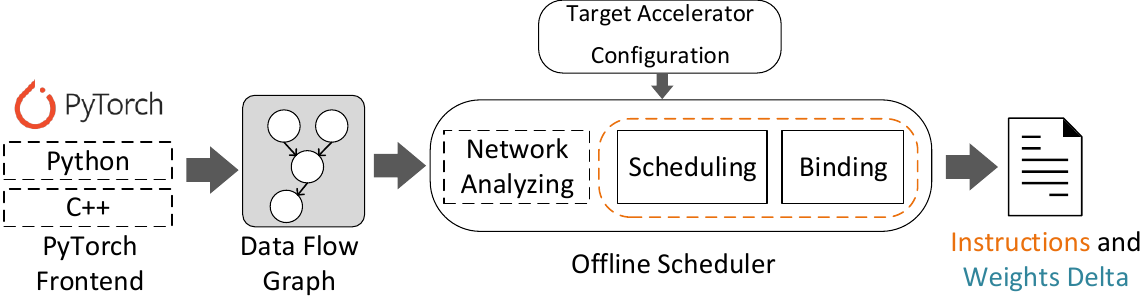}
    \vskip -0.15in
    \caption{ARAS Offline Flow.}
    \label{fig:Compilation_Flow}
    \vskip -0.15in
\end{figure}

\subsection{Baseline Scheduler}\label{subs:Scheduler}
Figure~\ref{fig:Compilation_Flow} shows an overview of the offline flow to configure ARAS. Initially, the Data Flow Graph is extracted from the PyTorch model, which includes the network architecture along with the execution order of the layers and their respective weights. In the subsequent stage, the scheduler analyzes the network architecture, receives the configuration of the accelerator, and calculates the number of PEs and APUs that are required to execute each layer. Next, it runs two different procedures to perform the scheduling and binding of tasks simultaneously. Finally, the ARAS offline flow provides a sequence of instructions for both updating weights, including the deltas, and computing the operations of each layer.

\begin{figure}[t!]
    \centering
    \includegraphics[width=0.85\columnwidth]{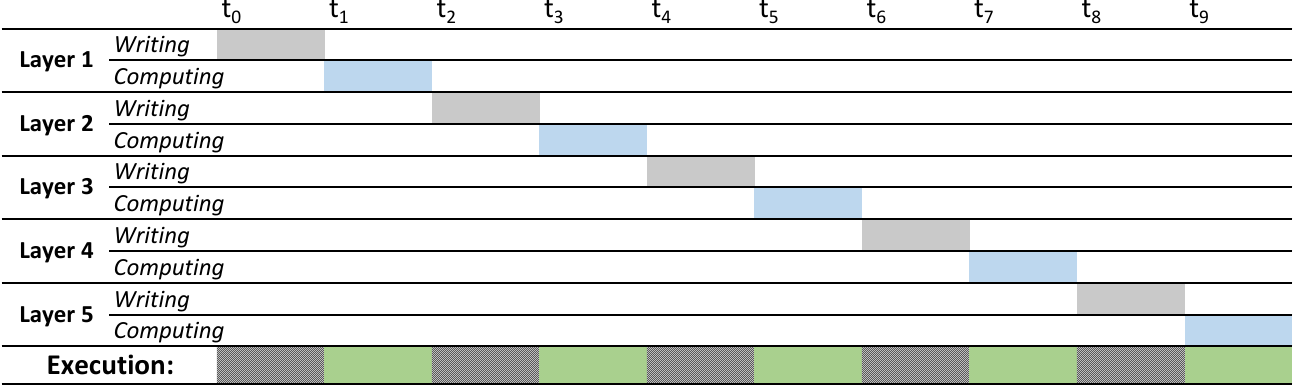}
    \vskip -0.10in
    \caption{Naive Scheduler.}
    \label{fig:Basic_scheduler}
    \vskip -0.15in
\end{figure}

Figure~\ref{fig:Basic_scheduler} introduces a simple naive scheduler that follows a sequential layer-by-layer execution pattern. Initially, all weights of a particular layer are written into the ReRAM crossbars and, then, dot-product operations are carried out for that layer. This sequence continues until the last layer is processed. In the event that a layer exceeds the capacity of the accelerator, it is divided into smaller segments, each of which is executed sequentially. However, this naive scheduler has several limitations as it cannot efficiently utilize all the available resources, particularly when dealing with small layers that only occupy a fraction of the resources. Furthermore, this approach is unable to hide the write latency, since writing a layer's weights is done in series with its computations.


\begin{figure}[t!]
    \centering
    \includegraphics[width=0.80\columnwidth]{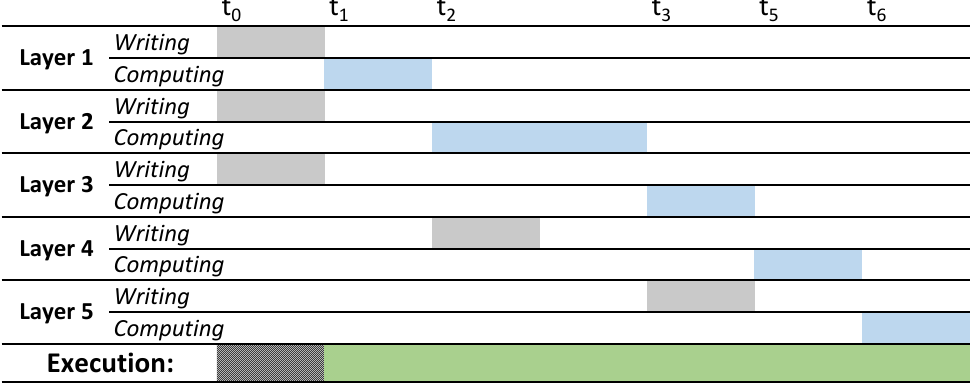}
    \vskip -0.10in
    \caption{ARAS Baseline Scheduler.}
    \label{fig:Baseline_Scheduler}
    \vskip -0.20in
\end{figure}

Conversely, we propose a more sophisticated offline scheduler that has the capability to overlap the latency of writing weights with the computations. Our scheduler statically fixes the accelerator resource allocation along with the layers' execution order due to the deterministic behavior of DNN inference, avoiding the high-cost of doing it online. Figure~\ref{fig:Baseline_Scheduler} shows an example of the ARAS baseline scheduler. The ARAS accelerator still conducts computations of one layer at a time due to data dependencies between layers, but our scheduler intelligently determines when to initiate the writing of layers' weights based on the availability of free resources. Upon completion of a layer's computation, the accelerator releases its resources, and the scheduler assigns those resources for the writing of the next layers. Based on the amount of crossbars, the accelerator can write the weights of a portion of a layer, an entire layer, or several layers, simultaneously. We can observe that the execution latency of the baseline scheduler is reduced in comparison to the naive scheduler. The main reason is due to the efficient overlap of the writing and computation phases.

The proposed scheduler takes decisions on how to manage and execute two distinct procedures that operate concurrently. The first procedure is responsible for delivering activations and configuring the components needed for neural computation. We refer to this as the "Computing Scheduling Procedure". In parallel, there is the "Weight Writing Scheduling Procedure", which handles the retrieval of weights, delta calculation, and mapping into ReRAM crossbars. Both procedures are coordinated and synchronized under certain conditions. The scheduler output comprises a list of instructions with a specific order, outlining the configuration of ARAS hardware components for doing dot-product operations and writing weights on the ReRAM crossbars. This order adheres to a layer-by-layer computation that starts the writing of new weights as soon as APUs become available. Next sections elaborate on the description of each procedure.

\subsection{Computing Scheduling Procedure}\label{subs:Computing_Procedure}
\begin{figure}[t!]
    \centering
    \includegraphics[width=1.0\columnwidth]{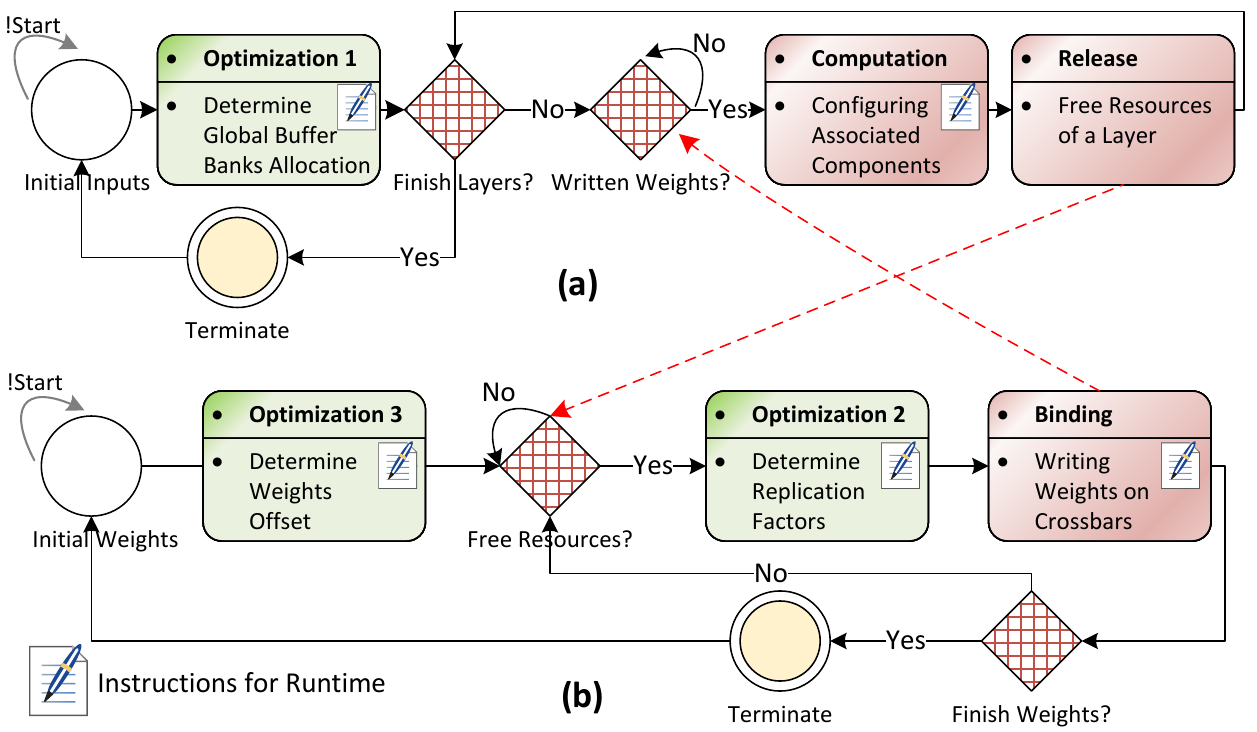}
    \vskip -0.10in
    \caption{ARAS Computing (a) and Weight Writing (b) Scheduling Procedures.}
    \label{fig:Schedure_Procedure}
    \vskip -0.15in
\end{figure}

Figure~\ref{fig:Schedure_Procedure}(a) outlines the computing procedure of the ARAS scheduler, which comprises five distinct states. The procedure starts with an initialization state triggered by a user request on a new DNN model or accelerator configuration. In this state, the scheduler produces instructions to transfer all network initial inputs from main memory to the Gbuffer. In the next state, the scheduler employs a heuristic to determine the ideal allocation of banks from the Gbuffer for each layer, an optimization discussed later in section~\ref{subs:Adaptive_Bank_Power_Gating}, modifying the configuration of the ARAS controller to store the activations in the corresponding banks. In the subsequent state, ARAS schedules the computations generating the instructions for dot-product operations layer-by-layer, waiting for the binding of new weights if necessary. Afterward, in the Release state, APUs that finished their assigned computations become free and ARAS starts binding new weights to the released APUs. Note that the Computation and Release states for a given layer are done back-to-back. Finally, after executing all layers the procedure ends, and the produced instructions can be used for multiple executions of the requested DNN model, restarting the procedure every time the model or hardware changes.


\subsection{Weight Writing Scheduling Procedure}\label{subs:Writing_Procedure}
Concurrently, the Weight Writing Scheduling Procedure, shown in Figure~\ref{fig:Schedure_Procedure}(b), starts along the computing procedure. First, the scheduler calculates the number of PEs and APUs needed for each layer by considering the layer's topology. In previous approaches~\cite{ISAAC, MNEMOSENE}, it was common to assign only the weights of one layer to each Tile/PE due to the complexity of managing multiple layers in a single Tile/PE. In contrast, our scheduler can efficiently accommodate different layers of weights within a single PE, yet each row of APUs in a PE belongs to one layer, since the same activations are broadcast to all APUs in the row. Next, ARAS employs an optimization, introduced in section~\ref{subs:Weight_Reuse}, to increase partial weight reuse and reduce energy consumption of ReRAM writing.


Then, the scheduler evaluates the availability of resources. In the following state, ARAS benefits from an optimization, described in section~\ref{subs:Adaptive_Replication}, that determines a set of candidate layers to write, together with their replication factors, according to the available resources. In the binding state, the scheduler produces instructions for writing the candidate layers and assigns the corresponding resources by binding the APUs. ARAS optimally seizes the opportunity to write as many weights as possible whenever there are free resources, which may lead to writing only part of a layer when resources are insufficient to accommodate all of its weights. The procedure concludes when there are no more weights for which to plan their writing. Otherwise, the procedure waits to have available resources to continue. Similar to the computing procedure, the writing procedure starts again when a new request arrives.







\section{ARAS Optimizations}\label{s:Optimizations}


\subsection{Adaptive Bank Selection}\label{subs:Adaptive_Bank_Power_Gating}
The Global Buffer (Gbuffer) in Figure~\ref{fig:Top_view} is in charge of storing the network's input and intermediate activations. Our Gbuffer is sized to accommodate one layer at a time, the one computing, fitting all input and output activations from any layer in our set of DNNs. However, if there is a layer that is larger, ARAS can partition the activations to process them in multiple iterations, similar to when not all the weights of a layer fit in the accelerator crossbars. Typically, the Gbuffer is able to accommodate the biggest layer, so only a fraction is actively used for most of the small layers. This underutilized storage results in high static power consumption.


To address this issue, we employ a multi-bank storage technique, a strategy commonly leveraged in advanced systems. Unlike previous approaches that employed a group of homogeneous banks of the same size, our Gbuffer includes banks of varying sizes. For each layer, the scheduler identifies the optimal set of banks to accommodate the input and output activations. Consequently, during the computation of a layer, only the banks required to store its activations are enabled, while the unused banks are power-gated. This approach ensures that the smallest set of banks is selected for each layer, minimizing static power consumption.


We formulate the process of identifying the optimal combination of banks for each layer as an Integer Linear Programming (ILP) problem with the objective of minimizing the Gbuffer static power during the execution of each layer. Equation~\ref{eqn:Goal} shows how to compute the static power of the Gbuffer for the execution of a layer \(l\), where \(a_i\) is a boolean variable indicating if the \(ith\) bank is selected to be active or not, and \(LeakagePower_i\) corresponds to the static power of the \(ith\) bank. The main goal is to minimize \(StaticPower_{gbuffer}(l)\). In addition, Equation~\ref{eqn:Storage} shows the on-chip memory required for storing the activations of a layer, where \(Mem_i\) is the storage capacity of the \(ith\) bank. Note that the banks being used for storing the input activations can not be selected for storing the output activations of any given layer, hence, by introducing specific constraints, the ILP solver is guided to select the optimal banks.

\vskip -0.20in
\begin{equation}
\label{eqn:Goal}
\text{\footnotesize $StaticPower_{gbuffer}(l) = \sum_{i=1}^{K} (a_i \times LeakagePower_i)$}
\end{equation}

\vskip -0.10in
\begin{equation}
\label{eqn:Storage}
\text{\footnotesize $RequiredMeM(l) \leq \sum_{i=1}^{K} (a_i \times Mem_i)$}
\end{equation}

This optimization takes place after the initial state of the computing scheduling procedure. Therefore, once this optimization is completed, the scheduler determines the smallest set of banks for each layer. As a result, at runtime, the ARAS controller enables the selected banks for each layer, efficiently power-gating the unused banks and reducing static power.

\subsection{Adaptive Replication Scheme}\label{subs:Adaptive_Replication}
In Section~\ref{subs:Replication} we described how the replication of weights improves the throughput of a convolution layer by simultaneously processing multiple activation windows. Earlier studies~\cite{RAELLA, ISAAC} employed a 'greedy' replication approach aiming to make the best use of all PEs while balancing the execution flow. These studies assumed that all weights were pre-loaded onto the accelerator, overlooking the latency and energy costs of writing into additional ReRAM crossbars at runtime.


In contrast, ARAS presents an innovative approach introducing a resource-constrained accelerator that updates the ReRAM crossbars for every layer, and includes an adaptive replication scheme. The ARAS scheduler decides when/how to replicate a layer's weights based on its criticality, in terms of compute latency, and the availability of APUs, ensuring that any added latency and energy from replicating weights does not penalize the overall improvements.

Shallow layers in Convolutional Neural Networks (CNNs) often have large input feature maps (IFMs) and multiple activation windows, which leads to long computation time. However, these layers tend to have a small set of weights, making them prime candidates for weight replication to significantly reduce computation time. On the other hand, replication improves computation speed at the expense of a greater number of writes and occupied APUs. Consequently, a high replication factor may delay the writing of weights for subsequent layers, decreasing concurrent writings due to the limited availability of APUs, and harming total execution time and energy.

The proposed scheduler of ARAS determines the replication factor for each layer during the third state of the Weight Writing Scheduling Procedure. In this state, ARAS plans the writing order of the weights of each layer taking into account the free rows of APUs in any PE at any given time. The scheduler is also responsible for deciding which layers can be written with their respective replication factors. Algorithm~\ref{alg:Replication_Scheme} presents the replication scheme employed in ARAS. It is important to note that in our accelerator all APUs within a PE's row are exclusively allocated to store weights of a specific layer, and they collectively process the same input segments but weights of different kernels. If the available resources fall short of meeting the requirements for writing the weights of the subsequent layer $L$, only a partial portion of that layer is written without any replication (lines 2-3). Conversely, when the resources are insufficient to accommodate the concurrent writing of two subsequent layers $L$ and $L+1$, the scheme tries to replicate layer $L$ (lines 4-5), since only the weights of layer $L$ can be entirely written and maybe replicated.


\begin{algorithm}[t!]
\scriptsize 
\caption{Weight Replication Scheme}
\label{alg:Replication_Scheme}
\begin{algorithmic}[1]
    \Procedure{Replication}{$\#Free Resources, L$}
    \Statex {\hspace{2.75em} \textcolor{red}{// ComL(i): Return computation latency of the ith layer.}}
    \Statex {\hspace{2.75em} \textcolor{red}{// Resources(i): Return number of required resources for ith layer.}}    
    \If{\(\#Free Resources < Resources(L)\)}
        \State {$Skip\:Replication$}
    \ElsIf{\(\#Free Resources < Resources(L) + Resources(L+1)\)}
        \State {$Replication(L, \#Free Resources)$}
    \Else
        \Statex {\hspace{2.75em} \textcolor{red}{// Accelerator has enough space for 2 layers or more.}}
        \Statex {\hspace{2.75em} \textcolor{red}{// Check how many layers can be fitted without replication.}}
        \State{$K = GetNumberOfLayers(L, \#Free Resources)$}
        \While{$!finish$}
            \State {$\#Free Resources \:+= ReleaseResources(L+K-1)$}
            \State {$ReplicateLongestLayers(\#Free Resources)$}
            \If{\(ComL(L+1) + ...+ ComL(L+K-2) \leq WL\)}
                \State{$finish = True$}
            \Else
                \State $K \gets K-1$
        \EndIf
    \EndWhile
    \EndIf
    \EndProcedure
\end{algorithmic}
\end{algorithm}

On the other hand, the proposed scheme adopts an iterative approach (lines 6-17) when plenty of rows of APUs within each PE are free. First, the scheduler aims to identify the number of consecutive layers $K$ that can be effectively written into the accelerator without replication, and then loops to find the best replication factors. The objective is to maximize both the amount of concurrently written layers and the overall improvement in total execution time. Within each iteration, the algorithm identifies the slowest layers and endeavors to enhance its performance by introducing replication. This is achieved by delaying the execution of the last layer $L+K-1$ and reallocating its resources to bolster the slowest layers' processing. The freed resources are allocated by a function (line 10) that iteratively allocates a portion of the available resources to the slowest layer in each iteration and continues until all available resources are exhausted. This procedure is repeated until the computation latency of all the candidate layers, except layer $L$, is reduced to a point where further replication no longer contributes to higher performance. This inflection point is determined by ensuring that the computation latency does not exceed the writing latency $WL$ of the following layers, which start to write after finishing the computations of layer $L$ and releasing its resources, as exceeding this threshold offers no further gains but penalties. Note that the writing latency remains relatively constant, whether it involves writing to a single APU or to multiple APUs simultaneously, due to the concurrent nature of the ReRAM writing process.


\begin{figure}[t!]
    \centering
    \includegraphics[width=0.60\columnwidth]{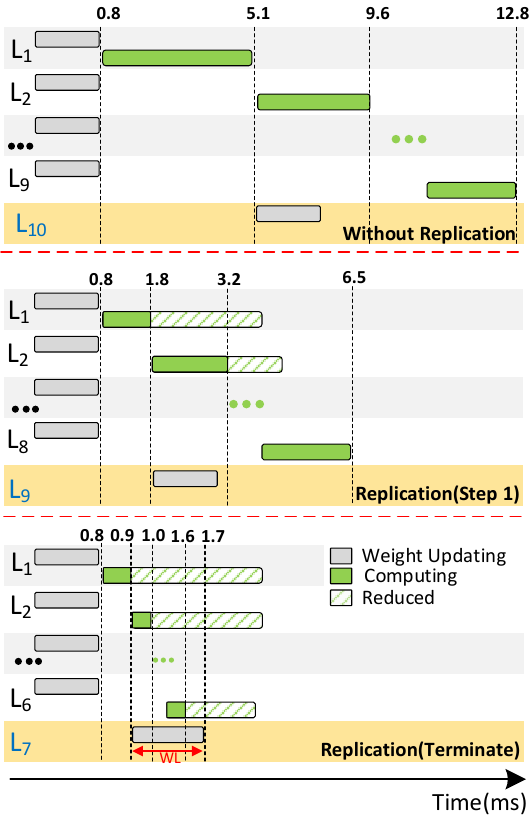}
    \vskip -0.10in
    \caption{An example of the replication scheme for the VGG16 network.}
    \vskip -0.20in
    \label{fig:Replication_Diagram_flow}
\end{figure}

Figure~\ref{fig:Replication_Diagram_flow} illustrates the ARAS replication scheme in action. In particular, this example shows the replication strategy for VGG16. At the beginning, the accelerator has the capacity to allocate resources for writing the weights of the first 9 layers without any replication. In the first iteration, the algorithm opts to delay the execution of $L_9$ and redistributes its resources to enable replication for other layers. This process is repeated for two additional iterations, after which a key observation arises: as the computation latency from $L_2$ to $L_6$ falls below the writing latency of the next layers $WL$, further replication would not yield performance improvements. This is because the accelerator must await the completion of weight writing for $L_7$ before processing more layers due to the layer-by-layer computation. Thus, at this point ARAS achieves the optimal resource utilization and performance. As illustrated in Figure~\ref{fig:Schedure_Procedure}, the weight replication scheme is triggered by the scheduler whenever there are available resources, determining the number of concurrent layers that can write next and their respective replication factors. In this particular example, upon completing the computations of $L_1$, the resources allocated to it are freed, prompting the replication scheme to initiate its operation for the following layers.

\subsection{Adaptive Partial Weight Reuse}\label{subs:Weight_Reuse}
As discussed in Section~\ref{s:intro}, ReRAM cells exhibit high energy consumption for writing, an inherent characteristic that may offset the advantages of employing ReRAM crossbars as a computational engine. Consequently, the creation of a scheme focused on alleviating the substantial energy costs associated with weight writing to ReRAM cells has the potential to make ReRAM-based DNN accelerators more appealing in comparison to established commercial platforms like TPU.

Expanding upon the details outlined in Section~\ref{subs:ReRAM}, the process of updating a multi-level ReRAM cell, rather than writing an absolute value, involves incrementing or decrementing its current value to reach the desired final value. Consequently, the energy consumption associated with updating ReRAM cells is strongly correlated with the magnitude of the delta between the cell's current value and its next value. The ideal scenario occurs when the current value of the cell matches its subsequent value, skipping the need for cell updates and saving the corresponding energy. In order to further reduce the energy consumption during the writing process, ARAS adopts a strategy that emphasizes increasing the equality or similarity of weights across layers that overwrite each other into the same crossbars cells. This approach aims to minimize the deltas between the values being written in the same ReRAM cells, thereby mitigating the energy overhead of updating these cells.

\begin{figure}[t!]
    \centering
    \includegraphics[width=0.55\columnwidth]{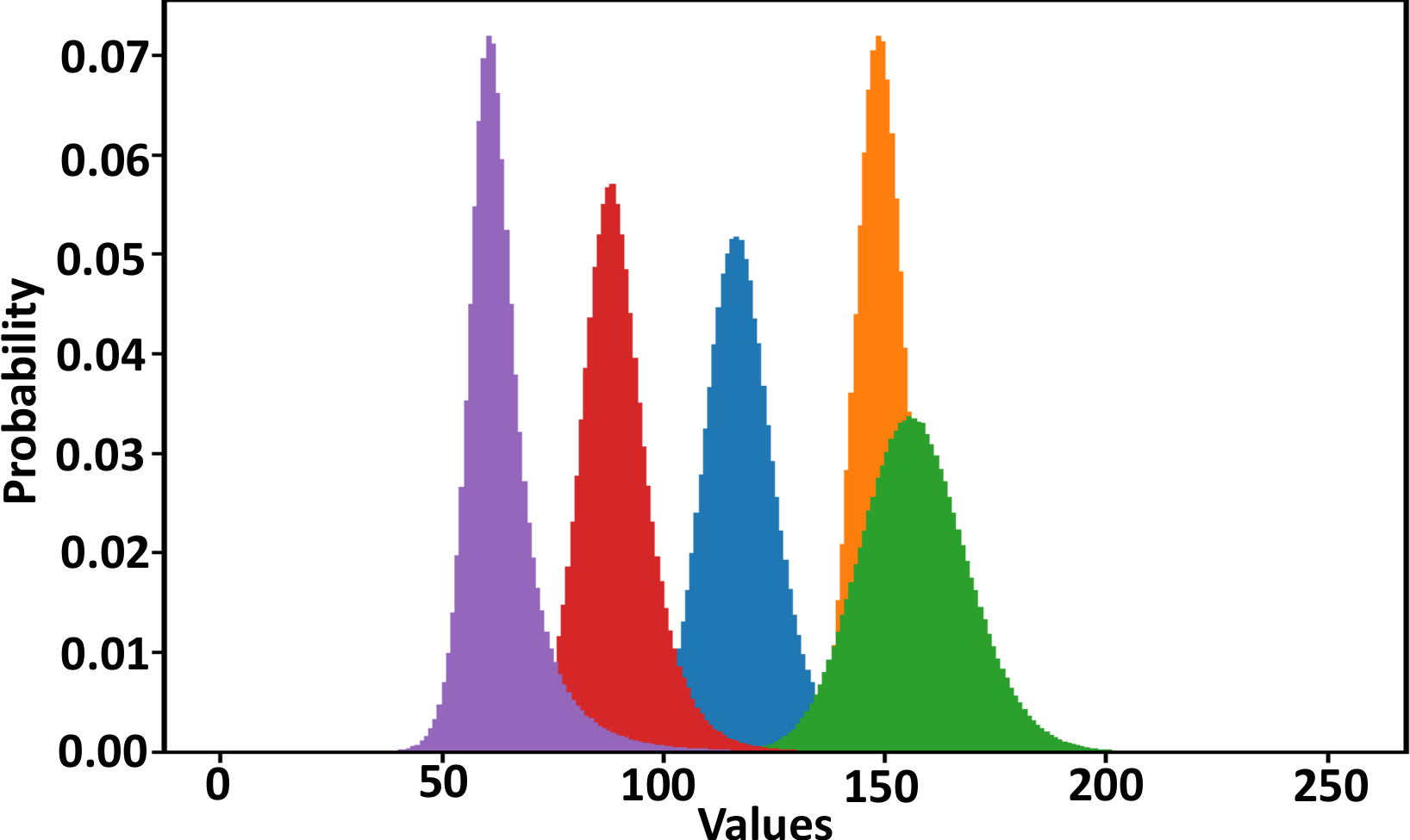}
    \vskip -0.10in
    \caption{Histogram of quantized weights for the last 5 layers of ResNet-50.}
    \label{fig:Quantized_weights_histogram}
    \vskip -0.15in
\end{figure}

Due to the inherent characteristics of ReRAM~\cite{multibit} cells, most ReRAM-based accelerators store each weight using multiple cells. In this work, activations and weights are both quantized into 8-bit integers (INT8), while ARAS uses a precision of 2 bits per ReRAM cell. Consequently, four cells are required to store a single weight. In Figure~\ref{fig:Quantized_weights_histogram}, we present the histogram of weights of a set of consecutive layers from ResNet-50 after applying uniform INT8 quantization. We can observe that the weight distribution across these layers exhibits distinct patterns. This variance diminishes the likelihood of achieving similarity when overwriting the weights of the preceding layer(s), particularly within the cells responsible for storing the Most Significant Bits (MSB) as described below.

\begin{figure}[t!]
    \centering
    \includegraphics[width=0.75\columnwidth]{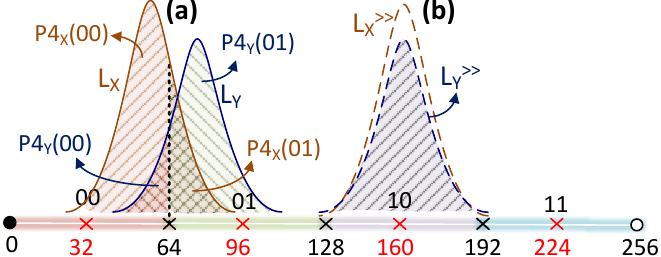}
    \vskip -0.10in
    \caption{Example of the shifting of weights of two layers.}
    \label{fig:Weight_shifting}
    \vskip -0.20in
\end{figure}

Figure~\ref{fig:Weight_shifting} depicts the distribution of the original quantized weights (a) for two given layers and an example of shifting (b) those weights to the right. After executing layer \(L_X\), the associated ReRAM cells are overwritten by the weights of layer \(L_Y\). The horizontal axis of this figure is divided into four sections with different colors, each representing a different bitwise value in the 4th cell, which contains the two most significant bits of each weight. Furthermore, \(Pi_X(k)\) (\(Pi_Y(k)\)) shows the probability that the ith cell for each weight of \(L_X\) (\(L_Y\)) has a value of \(k\), where \(k\) can be one of $00, 01, 10, 11$.

Equation~\ref{eq:Similarity} quantifies the likelihood that the ith ReRAM cell retains the same value across layers \(L_X\) and \(L_Y\). This probability directly correlates with the ratio of cell updates that can be skipped during the weight writing process, resulting in energy savings for the ReRAM cells. Note that the accumulation of \(Pi_X(k)\) for the four possible values is one.

\vskip -0.15in
\begin{equation}
\begin{footnotesize}
\begin{multlined}
\label{eq:Similarity}
Sim(X, Y, i) = Pi_X(00) \times Pi_Y(00) + Pi_X(01) \times Pi_Y(01) \\ \!  
+ Pi_X(10) \times Pi_Y(10) + Pi_X(11) \times Pi_Y(11) 
\end{multlined}
\end{footnotesize}
\end{equation}

In modern DNNs, it is common to observe that the four least significant bits of layers' weights, which correspond to the 1st and 2nd ReRAM cells, exhibit a nearly uniform distribution~\cite{RAELLA}. As a result, according to Equation~\ref{eq:Similarity}, the skipping ratio for these two cells is approximately $0.25$. In contrast, the other two ReRAM cells, responsible for storing the most significant bits, present a probability distribution more concentrated in a particular value of \(k\), influencing the skipping ratio of the cell. Based on Equation~\ref{eq:Similarity}, the ideal skipping ratio for the 3rd and 4th cells is attained when \(Pi_X(k)\) and \(Pi_Y(k)\) both reach their maximum probability at the same value of \(k\). Consequently, adjusting the weights of both layers to align the peaks of \(Pi_X\) and \(Pi_Y\) to the same value \(k\) can lead to an increased skipping ratio for these cells.


Given the bell-shaped distribution of weights in DNNs, the majority of weights in a layer tend to cluster around the mean. In order to attain a high skipping ratio for the 4th cell, we shift the mean values of weights in both \(L_X\) and \(L_Y\) to the midpoint of a specific section, as depicted in red in Figure~\ref{fig:Weight_shifting}(b). This approach ensures that the highest possible number of weights in both layers share the same value \(k\) in the 4th cell, thereby maximizing both \(P4_X\) and \(P4_Y\), and consequently $Sim(X, Y, 4)$. To optimize the skipping ratio for the 3rd cell, we can adjust the mean point of both layers in a manner that maximizes both \(P3_X\) and \(P3_Y\) for a value of \(k\). Similar to the strategy used for the 4th cell, if each section of Figure~\ref{fig:Weight_shifting} is further subdivided into four equal parts, shifting the mean of both layers to the middle point of one of these parts maximizes \(P3_X\) and \(P3_Y\), leading to an increased skipping ratio for the 3rd cell. However, the positions that maximize the skipping ratio for the 3rd and 4th ReRAM cells are distinct, making it difficult to maximize both simultaneously.


Equations \ref{eq:Shifting} and \ref{eq:Offset} present the shifting formula, where \(Weights(l)\) is the weight matrix of a given layer $l$, \(Offset\) is the value by which all weights within that layer are adjusted, and \(Center\) denotes the common mean value determined by the scheduler. Since the shifted weights are subject to the same numerical precision constraints, weights that after the shifting process fall outside the defined numerical precision ranges are clipped to the respective boundaries of these ranges. The clipping operation may result in accuracy loss, hence, the main objective of ARAS is to identify a specific \(Center\) where the weights exhibit maximum similarity while incurring in negligible accuracy loss.

\vskip -0.15in
\begin{equation}
\label{eq:Shifting}
\text{\footnotesize $ NewWeights(l) = Clip(Weights(l) +Offset \:,\:0\:,\:2^{n} - 1 )$}
\end{equation}

\vskip -0.15in
\begin{equation}
\label{eq:Offset}
 \text{\footnotesize $Offset = round(Center-mean(Weights(l)))$}
\end{equation}

In total, there are 20 different centers with the potential to maximize the combined skipping ratio of the 3rd and 4th cells. However, many of these centers (e.g. 32, 224) may lead to a significant accuracy loss due to a high clipping rate. To address this, ARAS disregards those centers near the limits of the range, and retains only six potential centers, namely (88, 104, 96, 160, 152, 168), where there is a chance of maximizing the combined skipping ratio for the 3rd and 4th cells with negligible accuracy loss. ARAS systematically evaluates all six centers, selecting the one that incurs the lowest accuracy loss and achieves the highest skipping ratio. Typically, introducing inaccuracies in the first layer of a DNN can lead to a substantial loss in model accuracy~\cite{han2015learning}. Therefore, ARAS avoids shifting the weights in the first layer and instead relies on the use of the original weights.

The dot-product results must be recomputed before transferring them to the next layer in order to account for the effect of the weight-shifting procedure. Equation~\ref{eq:Quantizing} shows the quantization function, where \(W_q\) and \(W_f\) are the quantized and floating-point representations of a given weight respectively, and \(q_w\) and \(zp_w\) denote the scaling factor and zero point, which are two common parameters of uniform quantization. In addition, Equation~\ref{eq:MVM} illustrates the process of the dot-product operation, where \(y_f\) is the FP representation of the dot-product result and \(i\) iterates on the number of vector elements. In order to remove the impact of adding the \(Offset\) to the quantized weights, the same \(Offset\) value is subtracted from \(zp_w\). In most DNN accelerators, the output feature maps (results of dot-product operations) are de-quantized before being transferred to the next layer. ARAS effectively compensates the shifting process by adjusting \(zp_w\) in the de-quantization procedure without introducing any additional overhead for recomputing the original dot-product.

\vskip -0.05in
\begin{equation}
\label{eq:Quantizing}
\text{\footnotesize $W_q = round(q_w \times W_f - zp_w)$}
\end{equation}

\vskip -0.15in
\begin{equation}
\label{eq:MVM}
 \text{\footnotesize $y_f = \sum_{i} \dfrac{x_q+zp_x}{q_x} \times \dfrac{(w_q+ \textcolor{green}{Offset})+\overbrace{(zp_w- \textcolor{red}{Offset})}^{Adjusted\:zp_w}}{q_w} + b_f$}
\end{equation}

\section{Methodology}\label{s:Methodology}
We have developed an event-driven simulator for ReRAM-based DNN accelerators that accurately models ARAS. Moreover, our ARAS scheduler assigns the resources and plans the execution order of a given DNN statically. Regarding area, latency, and energy consumption evaluation, the ReRAM crossbars are modeled using NeuroSim~\cite{Neurosim_github}, while on-chip buffers are characterized using CACTI-P~\cite{cacti-p}. On the other hand, the logic components are implemented in Verilog, and synthesized with Design Compiler~\cite{Design_compiler} employing the 28/32nm technology library from Synopsys. Regarding main memory, we model an LPDDR4 of 8 GB with a bandwidth of 19.2 GB/s (single channel) using DRAMSim3~\cite{DRAMsim3}.


To compare with a popular computational platform, we select Google's TPU, which, like ARAS, is an adaptive and efficient platform highly optimized for DNN execution. We have extended ScaleSim~\cite{ScaleSim} to model a TPU-like accelerator with the same tools as the ones used to model ARAS. ScaleSim is a simulator for modeling systolic array-based accelerators that employ a dataflow model and hardware components similar to those found in \cite{TPU, TPUV2, crew}. Table~\ref{tab:ParamTPU} shows the relevant parameters of the TPU-like accelerator that we use in this work. Total Data Buffer Size includes Activation Buffer and Weight Buffer together. To have a fair comparison, we scale the number of MAC units and the data buffer in the TPU-like accelerator to match the same area as ARAS ($26~{mm}^2$). Furthermore, both ARAS and TPU operate at the same frequency. This configuration ensures a reasonable baseline for evaluating ARAS's performance and energy efficiency.

\begin{table}[t!]
\caption{{Parameters of the TPU-like accelerator.}}
\label{tab:ParamTPU}
\centering
\resizebox{0.5\columnwidth}{!}{%
    \centering
    \begin{tabular}{|c|c|}
    \hline
    \cellcolor[gray]{0.9} {Technology} & {32 nm} \\
    \cellcolor[gray]{0.9} {Frequency} & {1 GHz} \\ 
    \cellcolor[gray]{0.9} {Number of MAC Units} & {64x64}\\
    \cellcolor[gray]{0.9} {Total Data Buffer Size} & {4.5 MB}\\
    \cellcolor[gray]{0.9} {Data Precision} & {8-bits}\\
    \hline
    \end{tabular}%
}
\vskip -0.20in
\end{table}

We model four different configurations of ARAS: a baseline \(ARAS\) without any optimization, \(ARAS_B\) with the Adaptive Bank Selection, \(ARAS_{BR}\) with both the first optimization and the Adaptive Replication Scheme, and finally \(ARAS_{BRW}\) with all the optimizations combined including the Partial Weight Reuse. In all cases, activations and weights are quantized to 8-bit integers without incurring any accuracy loss for our DNNs. Table~\ref{tab:Param} shows the parameters for all the configurations of ARAS. We set the number of PEs to 96, where each PE consists of 6x4 APUs, the crossbar size in each APU is 128x128 ReRAM cells, and each cell has a 2-bit resolution. Thus, 8-bit weights are represented with 4 consecutive cells. The Global Buffer is sized based on the worst-case scenario, which corresponds to the layer with the largest amount of input and output activations. In the baseline there are 15 banks, each with a size of 256kB, whereas the other configurations employ a heterogeneous bank scheme incorporating banks with the following sizes: 2x1KB, 2KB, 4KB, 64KB, 128KB, 256KB, 512KB, 1MB, and 2MB. All configurations have power-gating for the unused banks. Our optimizations do not require any additional hardware but slightly more on-chip memory with minor area overhead.



\begin{table}[t!]
\caption{Parameters for all the ARAS accelerator configurations.}
\label{tab:Param}
\centering
\resizebox{0.6\columnwidth}{!}{%
    \centering
    \begin{tabular}{|c|c|}
    \hline
    \cellcolor[gray]{0.9} Technology & 32 nm \\
    \cellcolor[gray]{0.9} Frequency & 1 GHz \\ 
    \cellcolor[gray]{0.9} Number of ADCs per APUs & 16 \\
    \cellcolor[gray]{0.9} ADCs Sampling Precision & 6-bits \\
    \cellcolor[gray]{0.9} PE Buffers Size & 1.5 KB \\
    \cellcolor[gray]{0.9} Crossbar Computation Latency & 96 Cycles \\
    \cellcolor[gray]{0.9} Crossbar Writing Latency & 768000 Cycles \\
    \hline
    \end{tabular}%
}
\vskip -0.20in
\end{table}



We evaluate our scheme on five state-of-the neural networks for image classification and Natural Language Processing (NLP) including VGG-16~\cite{VGG}, ResNet-50~\cite{ResNet}, DenseNet-161~\cite{DenseNet}, BERT-Base, and BERT-Large~\cite{BERT}. The ImageNet~\cite{ImageNet} dataset is used to evaluate the CNNs for image classification, while both BERT models are implemented by NVIDIA~\cite{NvidiaBERT}, trained on the Wikipedia dataset~\cite{wikidump}, and fine-tuned with SQuAD v1.1~\cite{SQuAD} for the question-answering task.

Among all the implemented optimizations, only the Partial Weight Reuse described in Section~\ref{subs:Weight_Reuse} may impact accuracy, but this impact is negligible. Table~\ref{tab:Accuracy_comparison} presents the accuracy loss, in absolute terms, after applying all optimizations. For CNNs, the accuracy is measured in terms of top-1 accuracy, while for BERT, it is measured using F1 accuracy. On average, ARAS has less than 0.12\% accuracy loss.

\begin{table}[t!]
\caption{Accuracy loss of ARAS for each DNN.}
\label{tab:Accuracy_comparison}
\centering
\resizebox{1.0\columnwidth}{!}{%
    \centering
    \begin{tabular}{|c|ccccc|}
    \hline
    \textbf{Network} & \textbf{VGG} & \textbf{ResNet} & \textbf{DenseNet} & \textbf{BERT-Base} & \textbf{BERT-Large}\\
    \hline
    \textbf{Accuracy loss} & 0\% & 0\% & 0.15\% & 0.02\% & 0.42\%\\
    \hline
    \end{tabular}%
}
\vskip -0.15in
\end{table}

\section{Experimental Results}\label{s:Results}
This section evaluates the performance, energy efficiency, and ReRAM writing activity of our proposal. First, we conduct an analysis of the total number of voltage pulses required to write into ReRAM crossbars after applying the weight reuse optimization. Second, we present the speedups achieved by ARAS due to the replication scheme. Then, we provide a breakdown of energy consumption and discuss the impact of each optimization. Next, we report the lifespan of ARAS. Finally, we compare the performance and energy of ARAS to that of a TPU-like accelerator.


\subsection{ReRAM Writing Activity Analysis}\label{subs:Weight_Reuse_results}
\begin{figure}[t!]
    \centering
    \includegraphics[width=0.9\columnwidth]{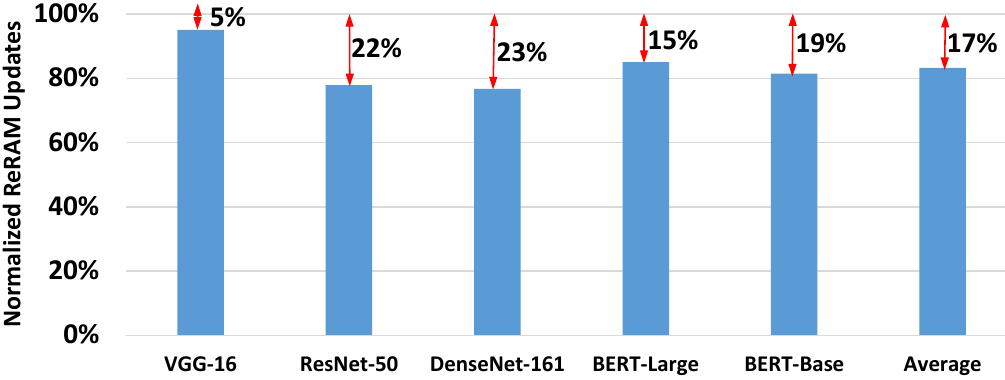}
    \vskip -0.10in
    \caption{Normalized ReRAM writing activity.}
    \label{fig:similarity}
    \vskip -0.20in
\end{figure}

Figure~\ref{fig:similarity} reports the normalized ReRAM writing activity, in terms of total voltage pulses required to update the cells, after the Partial Weight Reuse optimization. On average for our set of DNNs, \(ARAS_{BRW}\) reduces the total pulses by 17\% over the \(ARAS\) baseline. The important reduction of ReRAM writing activity is mainly due to an increase in the similarity of weights overwriting each other, not only because of the equal values but also a reduction in the magnitude of the weight deltas, requiring less pulses to update each ReRAM cell.


\subsection{ARAS Performance Evaluation}\label{subs:Performance}
\begin{figure}[t!]
    \centering
    \includegraphics[width=0.9\columnwidth]{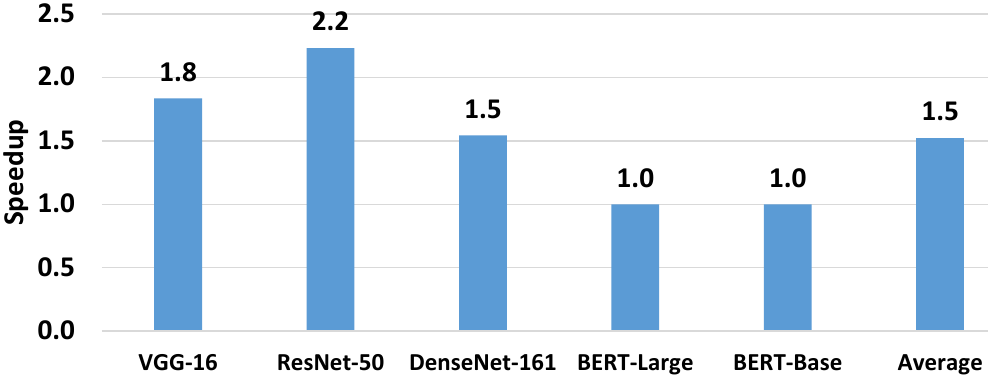}
    \vskip -0.10in
    \caption{Speedups of ARAS over the baseline without any optimization.}
    \label{fig:SpeedUp}
    \vskip -0.20in
\end{figure}

Figure~\ref{fig:SpeedUp} shows the speedups of \(ARAS_{BRW}\) over the baseline configuration without any optimization, achieving an average performance improvement of $1.5\times$. ARAS provides significant speedups for the three CNNs that range from $1.5\times$ (\textit{DenseNet-161}) to $2.2\times$ (\textit{ResNet-50}). The reduction in execution time is due to the efficient adaptive weight replication scheme of the CONV layers, described in Section~\ref{subs:Adaptive_Replication}. Therefore, FC-based DNNs, like BERT, do not see any benefit. The other optimizations do not target nor provide performance improvements. The main reason is that the worst case latency when writing the weights of a layer, and the most common, is to write in all the rows of a crossbar sequentially, requiring the highest amount of pulses in a cell of each row, even tough multiple crossbars can be written in parallel. Consequently, the partial weight reuse optimization requires to reduce the ReRAM writing activity of all the slowest cells of multiple rows of different crossbars at the same time, an effect that has not occurred in our set of DNNs.


In order to assess the effectiveness of the ARAS scheduler, we define an upper-bound of the performance as the inverse of the lowest execution time achievable per DNN, which is given by the total time required to write the weights of all the layers of a DNN model once, taking into account the amount of resources available. On average, the baseline performance is 66\% of the upper bound. In contrast, ARAS exhibits significantly higher performance, reaching 88\%, which is remarkably close to the upper-bound, and demonstrates the successful overlap between the computation and writing procedures.


Furthermore, the throughput of ARAS scales with the amount of PEs and APUs. Our current configuration (see Section~\ref{s:Methodology}) has been selected to achieve real-time performance in all DNNs of our set, while limiting the resources so that none of the models fit entirely in the ReRAM crossbars. In particular, the throughput of ARAS in terms of inferences per second (Inf/s) is: 43 (VGG-16), 132 (ResNet-50), 95 (DenseNet-161), 130 (BERT-Base, sequence length 128), and 39 (BERT-Large, sequence length 128).

\subsection{ARAS Energy Evaluation}\label{subs:Energy}
\begin{figure}[t!]
    \centering
    \includegraphics[width=1.0\columnwidth]{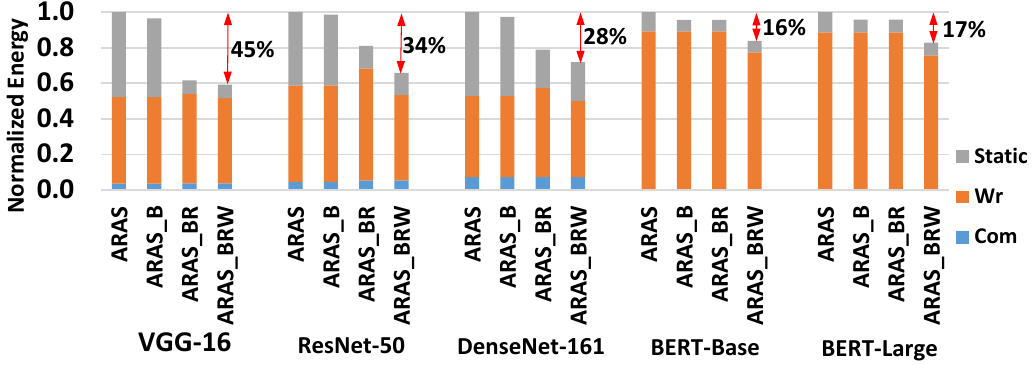}
    \vskip -0.10in
    \caption{Normalized energy breakdown of each ARAS configuration.}
    \label{fig:Energy_BreakDown}
    \vskip -0.20in
\end{figure}

Figure~\ref{fig:Energy_BreakDown} reports the energy breakdown of each configuration of ARAS normalized to the baseline without any optimization. We can observe that the computation energy is negligible compared to the writing and static energy. Specifically, NLP networks experience significant energy consumption due to weight writing, while in CNNs, static energy is quite high.

ARAS leverages two optimizations to reduce static energy, which is primarily consumed by the Gbuffer. First, the adaptive bank selection, discussed in Section~\ref{subs:Adaptive_Bank_Power_Gating}, reduces static energy by selecting the smallest set of banks for each layer to store input and output activations. On average, this optimization alone (\(ARAS_B)\) has a modest 3\% reduction in total energy. However, note that it does not incur any additional overhead. Furthermore, the adaptive weight replication scheme (\(ARAS_{BR})\) provides a reduction in static energy due to the performance improvements shown in Figure~\ref{fig:SpeedUp}. Although these benefits come at the expense of a small increase in writing energy, the last optimization offsets the cost. On average, the second optimization by itself contributes to a 14\% reduction in total energy.


In most modern DNNs, ReRAM writing dominates the total energy consumption. The adaptive partial weight reuse optimization in ARAS aims to reduce energy consumption when updating ReRAM cells with new weights. To achieve this, ARAS increases the bitwise similarity between the values stored in the ReRAM cells and the new values that will overwrite those cells, as described in Section~\ref{subs:Weight_Reuse_results}. On average, this optimization provides 11\% reduction in total energy consumption, which is well correlated with the reduction in ReRAM writing activity. Overall, \(ARAS_{BRW}\) achieves 28\% energy savings by leveraging all the three optimizations.

\subsection{ARAS Endurance Analysis}\label{subs:Lifespan}
As described in~\ref{subs:Writing}, there is a growing body of research demonstrating that the endurance of ReRAM cells can reach up to $10^{11}$ and $10^{12}$ writing cycles. Table~\ref{tab:lifetime} reports the lifespan (in years) for real-time inference execution with an endurance of $10^{11}$ cycles. In CNNs, real-time is defined as 30 Inf/s, while in BERT 100 Inf/s. Furthermore, it also shows the lifespan while the accelerator operates at maximum throughput (reported in~\ref{subs:Performance}) with an endurance of $10^{12}$ cycles. The results indicate that employing a server with the ARAS scheme can offer a minimum of 27 years of lifespan with an endurance of $10^{12}$ cycles. Besides, if the server is utilized for a diverse range of DNNs the lifespan can be extended to approximately 40 years (or more since this estimation assumes 100\% utilization). Consequently, the opportunities for ReRAM cells with enhanced endurance in the PUM domain are evident, prompting companies to explore new insulators such as TaOx. These insulators offer prolonged lifespan, which are crucial for harnessing ReRAM cells in computing applications, a direction encouraged by ARAS and other proposals\cite{On_Endurance}.

\begin{table}[t!]
\caption{ARAS lifespan in years for different DNNs.}
\label{tab:lifetime}
\centering
\resizebox{1.0\columnwidth}{!}{%
    \centering
    \begin{tabular}{|c|cccccc|}
    \hline
    \textbf{Network} & \textbf{VGG} & \textbf{ResNet} & \textbf{DenseNet} & \textbf{BERT-Base} & \textbf{BERT-Large}& \textbf{Average}\\
    \hline
    \textbf{Real-Time} $(10^{11})$ & 7 & 26 & 17 & 4 & 3 & 12\\
    \hline
    \textbf{Max Throughput} $(10^{12})$ & 50 & 27 & 56 & 28 & 27 & 40\\
    \hline
    \end{tabular}%
}
\vskip -0.15in
\end{table}

\subsection{Comparison With a TPU-Like Accelerator}\label{subs:TPU_like}
\begin{figure}[t!]
    \centering
    \includegraphics[width=1.0\columnwidth]{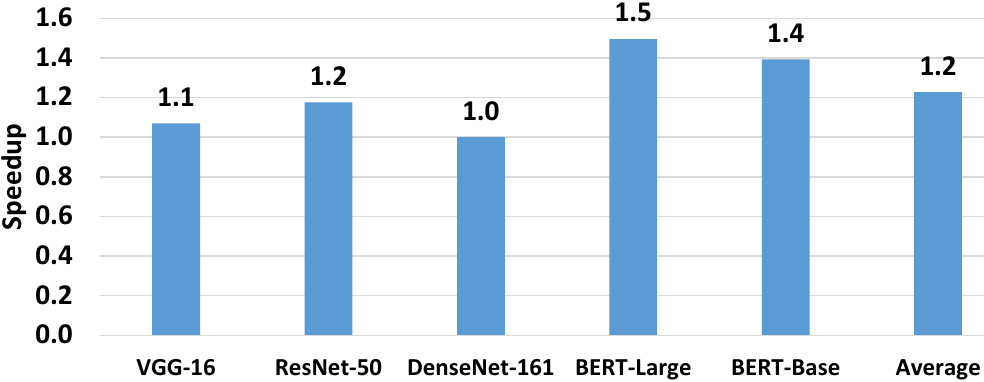}
    \vskip -0.10in
    \caption{Speedups of ARAS for each DNN over a TPU-like accelerator.}
    \label{fig:over_tpu_speedup}
    \vskip -0.15in
\end{figure}

\begin{figure}[t!]
    \centering
    \includegraphics[width=1.0\columnwidth]{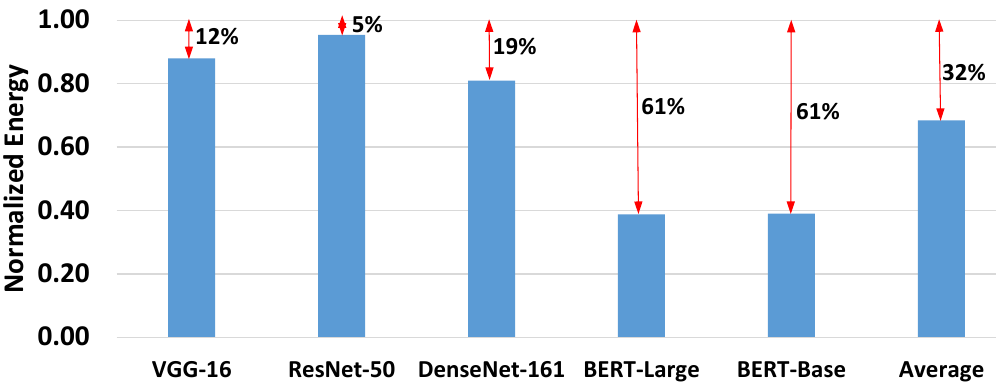}
    \vskip -0.10in
    \caption{Normalized energy for each DNN over a TPU-like accelerator.}
    \label{fig:over_tpu_energy}
    \vskip -0.20in
\end{figure}

We include a comparison of \(ARAS_{BRW}\) with a TPU-like accelerator in terms of performance and energy efficiency. For a fair comparison, TPU and ARAS have the same area and frequency. Figure~\ref{fig:over_tpu_speedup} shows the speedup of ARAS for the five DNNs, achieving an average performance improvement of $1.2\times$. Figure~\ref{fig:over_tpu_energy} reports normalized energy. On average, ARAS reduces the energy consumption by 33\%. ARAS's novel scheduler and its associated optimizations provide higher performance and energy efficiency over a TPU-like accelerator. Note that the moderate energy reduction observed in CNNs underscores the importance of our optimizations, proving that without them, the overhead of writing weights in the ReRAM crossbars could result in worse energy efficiency than the TPU. These results suggest that ARAS is a suitable candidate for DNN inference, effectively addressing the challenge of adaptability of previous ReRAM-based accelerators.

\subsection{ARAS Overheads}\label{sub:Overhead}
ARAS does not require any additional hardware but has a slight overhead in terms of the on-chip memory size and amount of written weights in ReRAM crossbars. For the Adaptive Bank Selection optimization, described in section~\ref{s:Methodology}, ARAS requires an additional 200 KB of Gbuffer memory due to the heterogeneous banks' sizes, resulting in about 5\% increase in the overall Gbuffer size. Moreover, the Adaptive Replication Scheme increases the total number of weights written in the ReRAM crossbars by an average of 6.4\% due to the replication of weights. This overhead is compensated by the significant energy savings and performance gains achieved through this optimization. Finally, as discussed in section~\ref{subs:Weight_Reuse}, the Adaptive Partial Weight Reuse optimization effectively compensates the shifting process by adjusting \(zp_w\) in the de-quantization procedure without introducing any additional overhead for recomputing the original dot-product. All ARAS optimizations are employed by our offline scheduler, so they do not incur any extra hardware to manage them at runtime, except for a few extra gates and signals needed to control the Adaptive Bank Selection. The area overhead of ARAS due to the additional on-chip memory is less than 1.5\%.


\section{Conclusions}\label{s:Conclusion}
In this paper, we show that previous ReRAM-based DNN accelerators face an \textit{adaptability} challenge due to the lack of flexibility and efficient mechanisms for updating weights in ReRAM crossbars at runtime. Then, we propose ARAS, a low-cost ReRAM-based accelerator that can be adapted to efficiently execute any present and future DNN, regardless of its model size, using limited resources. ARAS presents a novel scheduler that overlaps computations of a given layer with the writing of weights of subsequent layers. In addition, ARAS includes three optimizations aimed at mitigating the expensive overheads of writing weights in ReRAM. Our experimental results show that, on average, ARAS's smart scheduler and its optimizations provide $1.5\times$ speedup and 28\% reduction in energy consumption over the baseline without any optimization. Over a TPU-like accelerator, ARAS achieves $1.2\times$ speedup and 33\% reduction in energy.

\bibliographystyle{IEEEtranS.bst}
\bibliography{refs}

\end{document}